# Theoretical realization of rich magnon topology by symmetry-breaking in honeycomb bilayer ferromagnets

Doried Ghader

College of Engineering and Technology, American University of the Middle East, Eqaila, Kuwait

**Abstract.** We reveal the rich magnon topology in honeycomb bilayer ferromagnets (HBF) induced by the combined effect of interlayer exchange, Dzyaloshinskii-Moriya interaction (DMI), and electrostatic doping (ED). In particular, we present a systematic study of the Hamiltonian non-adiabatic evolution in the HBF parametric space, spanned by the symmetry-breaking terms (DMI and ED) and interlayer exchange. We determine the band closure manifolds which are found to divide the parametric space into six distinct regions, matched with five distinct topological phases and one topologically trivial phase. The characteristic Chern numbers and thermal Hall conductivities are calculated for the topological phases. Edge spectra, dictated by the bulk-edge correspondence, are also analyzed in the nanoribbon version of the model. Both bulk and edge spectra are found to be nonreciprocal as a consequence of ED and edge magnons are observed to counter propagate on opposite edges. The predicted results offer new insights on the manipulation of magnonic Chern numbers and magnon topological transport via experimentally tunable parameters.

## I. Introduction

Topological magnons [1-12] constitute an active research field in view of their technological potentials. In particular, topologically protected magnon boundary modes can be harnessed to realize backscattering-free magnonic waveguides, owing to their robustness against disorder and other variations in the sample. In this context, the rapidly growing field of 2D magnets [13-50] can offer novel opportunities for magnonic devices based on the 1D magnons confined to the edges or domain walls in 2D magnets.

Exploring new topological phases in 2D magnets is essential for further advancement towards their technological implementation in magnonics [51-53]. Topological magnons can only exist in magnetic materials with a gapped band structure. In an AB-stacked bilayer honeycomb ferromagnet (HBF), the in-plane Dzyaloshinskii-Moriya interaction (DMI) breaks time-reversal symmetry and opens topological gaps at the $\pm K$ valleys [17, 19, 54]. The DMI renders the HBF a Chern insulator with integer Chern numbers. The topological phase for this model is unique and the model Hamiltonian evolves adiabatically as a function of DM and interlayer interactions. Recently, HBF with layer dependent electrostatic doping (ED) [30] has been proposed to realize novel topological transport of valley-polarized magnons [44, 54]. The effect of layer dependent ED is to break the inversion symmetry which in turn gaps the magnonic spectrum and induces valley Chern numbers.

A formal analysis of the generalized model for HBF with both DMI and ED is still missing. Here, we study the consequences of coexisting DMI and ED on the magnon topology in HBF, predicting five distinct topological phases and one topologically trivial phase. Characteristic Chern numbers, thermal Hall conductivities, and nanoribbon edge spectra are calculated for each of the predicted topological phases. It is important to note that our recent study on valley-polarized magnons in HBF [54] briefly discusses some of these topological phases while the formal analysis was left to the present work.

## II. Momentum-space Hamiltonian

We consider an AB-stacked HBF in collinear ground state with nearest neighbor exchange interaction, DMI, and ED. For completeness, the geometry and choice of axes are presented in Fig. 1 (top view of the bilayer). We define the vectors $\vec{\delta}_i^A$ and $\vec{\gamma}_j$ which connect an A-site to its three nearest and six next nearest neighbors respectively. Vectors $\vec{\gamma}_j$ also serve the B-sublattice, whereas $\vec{\delta}_i^B = -\vec{\delta}_i^A$. The explicit form of these vectors are as follows: $\vec{\delta}_1^A = a(0, 1/\sqrt{3})$, $\vec{\delta}_2^A = a(1/2, -\sqrt{3}/6)$, $\vec{\delta}_3^A = a(-1/2, -\sqrt{3}/6)$, $\vec{\gamma}_1 = a(1/2, -\sqrt{3}/2)$, $\vec{\gamma}_2 = a(-1/2, -\sqrt{3}/2)$, $\vec{\gamma}_3 = a(1,0)$, $\vec{\gamma}_4 = -\vec{\gamma}_1$, $\vec{\gamma}_5 = -\vec{\gamma}_2$, and $\vec{\gamma}_6 = -\vec{\gamma}_3$. The lattice constant $a$ denotes the $A - A$ (or $B - B$) distance whereas the nearest neighbor distance is $a/\sqrt{3}$. We also define $\vec{\delta}_\perp$ connecting $A_1 - B_2$ dimers.

We adopt the semi-classical linear spin wave approach [55-64] to derive the momentum-space Hamiltonian. Holstein-Primakov approach yields identical results. The real space Hamiltonian can be expressed as

$$\mathcal{H} = -J \sum_{l, \vec{\delta}_i^A} \vec{S}^{A_l}(\vec{R}_{A_l}, t) \cdot \vec{S}^{B_l}(\vec{R}_{A_l} + \vec{\delta}_i^A, t) - J_\perp \sum_{\vec{R}_{A_1}} \vec{S}^{A_1}(\vec{R}_{A_1}, t) \cdot \vec{S}^{B_2}(\vec{R}_{A_1} + \vec{\delta}_\perp, t)$$
$$+ \sum_{\alpha, l, \vec{\gamma}_j} D_z(\vec{R}_{\alpha_l}, \vec{R}_{\alpha_l} + \vec{\gamma}_j) \vec{S}^{\alpha_l}(\vec{R}_{\alpha_l}, t) \cdot \vec{S}_D^{\alpha_l}(\vec{R}_{\alpha_l} + \vec{\gamma}_j, t) - \sum_{\alpha, l} U_l \hat{z} \cdot \vec{S}^{\alpha_l}(\vec{R}_{\alpha_l}, t)$$

(1)

The first, second and third terms in $\mathcal{H}$ account for the intralayer exchange, interlayer exchange and DM interactions respectively. The fourth term accounts for ED. $J$ and $J_\perp$ are the nearest neighbor in-plane and interlayer exchange coefficients respectively. The alternating next nearest neighbor DMI vector has the form $\vec{D}(\vec{r},\vec{r}+\vec{\gamma}_j) = D_z\hat{z} = \pm D\hat{z}$. The $\pm\hat{z}$ orientation of $\vec{D}$ is determined in the conventional way from the local geometry of the honeycomb lattice [8].

Index $l$ specifies the layer and is summed over 1 and 2. Index $\alpha$ denotes the sublattice and runs over A and B sites. $\vec{S}^{\alpha_l}(\vec{R}_{\alpha_l}, t)$ is the spin on site $\vec{R}_{\alpha_l}$ at time $t$. We have also introduced $\vec{S}_D^{\alpha_l} = S_y^{\alpha_l}\hat{x} - S_x^{\alpha_l}\hat{y}$ to express the DMI term in the form of a scalar-product rather than a cross-product [40, 42]. The ED potentials are denoted $U_l = \pm U$ for $l = 1, 2$ respectively.

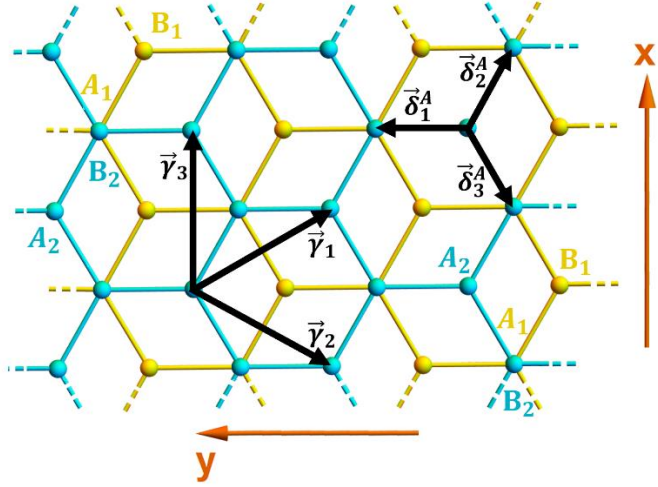

**Figure 1:** Schematic representation for a top view of the AB-stacked quasi-infinite honeycomb ferromagnet. The figure also illustrates the nearest and next nearest neighbors vectors.

The effective field acting on the $A_1$-sublattice can be deduced from $\mathcal{H}$ as [55-61, 63]

$$\vec{H}^{A_1}(\vec{R}_{A_1}, t) = -J_\perp \vec{M}^{B_2}(\vec{R}_{A_1} + \vec{\delta}_\perp, t) + U\hat{z} - J\sum_{\vec{\delta}_i^A} \vec{M}^{B_1}(\vec{R}_{A_1} + \vec{\delta}_i^A, t)$$

$$+ \sum_{\vec{\gamma}_j} D_z(\vec{R}_{A_1}, \vec{R}_{A_1} + \vec{\gamma}_j)\vec{M}_D^{A_1}(\vec{R}_{A_1} + \vec{\gamma}_j, t)$$

(2)

where $\vec{M}$ denotes the magnetization. In the semi-classical approach, spins are treated as numerical vectors and the spin dynamics are governed by the Landau-Lifshitz (LL) equations of motion, $\partial_t \vec{M}^{A_1} = \vec{M}^{A_1} \times \vec{H}^{A_1}$. The LL equations, keeping only linear terms, yield

$$i\partial_t M^{A_1}(\vec{R}_{A_1}, t) = (3JM_z + J_\perp M + U)M^{A_1}(\vec{R}_{A_1}, t) - JM_z \sum_{\vec{\delta}_i^A} M^{B_1}(\vec{R}_{A_1} + \vec{\delta}_i^A, t)$$

$$-J_\perp M_z M^{B_2}(\vec{R}_{A_1} + \vec{\delta}_\perp, t) - iM_z \sum_{\vec{\gamma}_j} D_z(\vec{R}_{A_1}, \vec{R}_{A_1} + \vec{\gamma}_j) M^{A_1}(\vec{R}_{A_1} + \vec{\gamma}_j, t)$$

(3)

with $M^{\alpha_l} = M_x^{\alpha_l} - iM_y^{\alpha_l}$. The symbol $M_z$ denotes the constant z component of the magnetization. For infinite bilayer, the translational symmetry is preserved along $x$ and $y$ directions, and the momenta (or wavenumbers) $k_x$ and $k_y$ are well defined. Fourier transformation can be implemented in both directions to arrive at the momentum-space equation of motion

$$i\partial_t M^{A_1}(\vec{k}, t) = [3JM_z + J_\perp M_z + U + DM_z f_D(\vec{k})]M^{A_1}(\vec{k}, t) - JM_z f(\vec{k})M^{B_1}(\vec{k}, t)$$
$$- J_\perp M_z M^{B_2}(\vec{k}, t)$$

(4)

with

$$f(\vec{k}) = e^{ik_y \frac{a}{\sqrt{3}}} + 2e^{-i\frac{\sqrt{3}a}{6} k_y} \cos\left(\frac{a}{2}k_x\right)$$

$$f_D(\vec{k}) = 4\sin\left(\frac{a}{2}k_x\right) \cos\left(\frac{\sqrt{3}a}{2} k_y\right) - 2\sin(k_x a)$$

Equations for $M^{B_1}$, $M^{A_2}$ and $M^{B_2}$ can be derived in a similar manner. Collecting the four momentum-space equations results in a Schrödinger matrix equation

$$i\partial_t |\Psi\rangle = \mathcal{H}(\vec{k})|\Psi\rangle \qquad (5)$$

with the 4-band momentum-space Hamiltonian

$$\mathcal{H}(\vec{k}) = JM_z \begin{pmatrix} \alpha_1 & -f(\vec{k}) & 0 & -v_0 \\ -f^*(\vec{k}) & \alpha_1 & 0 & 0 \\ 0 & 0 & \alpha_3 & -f(\vec{k}) \\ -v_0 & 0 & -f^*(\vec{k}) & \alpha_4 \end{pmatrix}$$

The parameters are $\alpha_1 = 3 + v_0 + U_0 + Df_D(\vec{k})/J$, $\alpha_2 = 3 + U_0 - Df_D(\vec{k})/J$, $\alpha_3 = 3 - U_0 + Df_D(\vec{k})/J$, $\alpha_4 = 3 + v_0 - U_0 - Df_D(\vec{k})/J$, $v_0 = J_\perp/J$ and $U_0 = U/(JM_z)$. In what follows, we will drop the multiplicative factor $JM_z$ from $\mathcal{H}(\vec{k})$ and work with energies normalized by $JM_z$.

### III. Bands topology

Diagonalizing $\mathcal{H}(\vec{k})$ yields 4 energy bands which we denote $[\epsilon_4, \epsilon_3, \epsilon_2, \epsilon_1]$ in descending energy order. For a gapped spectrum, the Berry curvature $B_n(\vec{k})$ for band $\epsilon_n$ can be calculated using

$$B_n = -Im \sum_{n' \neq n} \frac{[\langle n|\vec{\nabla}_{\vec{k}}\mathcal{H}|n'\rangle \times \langle n'|\vec{\nabla}_{\vec{k}}\mathcal{H}|n\rangle]_z}{(\epsilon_n - \epsilon_{n'})^2}$$

(6)

The numerator is the $z$-component of a cross-product, $\vec{\nabla}_{\vec{k}}$ is the momentum-space gradient, and $|n\rangle$ denotes the eigenfunction of $\epsilon_n$. The Chern number $C_n$ is next deduced as the integral of $B_n$ over the Brillouin zone (BZ)

$$C_n = \frac{1}{2\pi} \iint_{\vec{k} \in BZ} B_n(\vec{k}) dk_x dk_y$$

(7)

Finally, the thermal magnon Hall conductivity ($\kappa_{xy}$) for topological bands can be calculated numerically using the equation [17, 33, 42, 54, 65, 66],

$$\kappa_{xy} = -\frac{k_B^2 T}{\hbar V} \sum_{\vec{k},n} c_2\left(g\left(\epsilon_n(\vec{k})\right)\right) B_n(\vec{k})$$

(8)

The function $g(\epsilon_n) = [e^{\epsilon_n/k_B T} - 1]^{-1}$ represents the Bose-Einstein distribution, $c_1(x) = (1 + x)\ln(1 + x) - x\ln x$, $c_2(x) = (1 + x)\left[\ln\left(\frac{1+x}{x}\right)\right]^2 - (\ln x)^2 - 2\text{Li}_2(-x)$, $\text{Li}_2$ stands for the dilogarithm function, $V$ is the volume of the system and $k_B$ is the Boltzmann constant. We set $k_B = \hbar = 1$ in our numerical calculation of $\kappa_{xy}$.

The magnon spectra in HBF with either DMI or ED are characterized by protected valley band gaps that cannot be closed by varying the model parameters. Magnons in HBF with exchange and DMI are topologically nontrivial with a single topological phase [17, 19]. HBF with exchange and ED, however, is characterized by zero band Chern numbers over the complete BZ. Nevertheless, the model supports topological transport of valley-polarized magnons due to the nonzero valley Chern numbers [44, 54].

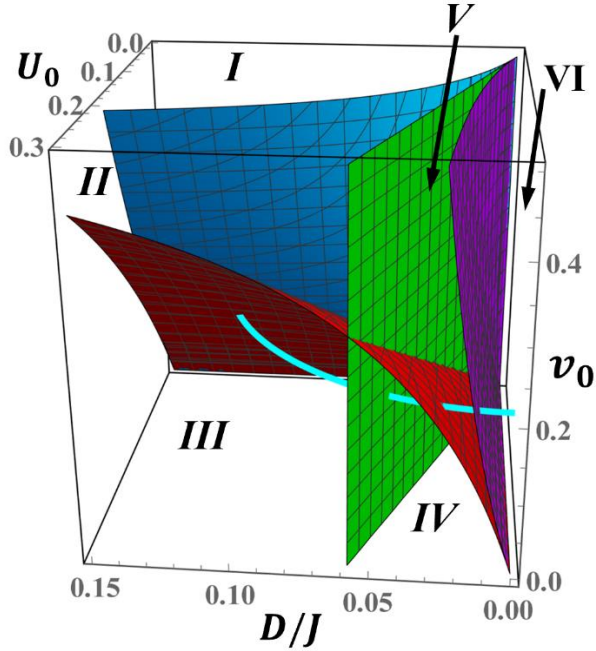

**Figure 2:** Numerically calculated band gap closure manifolds in the Hamiltonian parametric space. These manifolds divide the space into six regions denoted I, II, …, VI. Regions I, II, …, IV are topological whereas VI is topologically trivial. The cyan curve is an oriented elliptic arc penetrating all six phases of the HBF (parametric definition is given in the text).

The band topology becomes more exotic in the full $(v_0, U_0, D)$ parametric space, where the magnon spectrum of $\mathcal{H}(\vec{k})$ is gapped accept at specific manifolds in the space. Interestingly, while $U_0$ and $D$ both induce protected band gaps on their own, the interplay between $D$, $U_0$ and $v_0$ can close the gaps at $\pm K$. The Hamiltonian $\mathcal{H}(\vec{K})$ can be analytically diagonalized at the valley momentum $\vec{K} = (4\pi/3, 0)$, with four eigenvalues $3 + 3\sqrt{3}D/J - U_0$, $3 - 3\sqrt{3}D/J + U_0$, $3 +$ $v_0 - \sqrt{27(D/J)^2 + v_0^2 + 6\sqrt{3}DU_0/J + U_0^2}$, and $3 + v_0 + \sqrt{27(D/J)^2 + v_0^2 + 6\sqrt{3}DU_0/J + U_0^2}$. The order of these eigenvalues depends on the specific values of $D$, $U_0$ and $v_0$. At $-\vec{K}$, the eigenvalues have different expressions given by $3 - 3\sqrt{3}D/J - U_0$, $3 + 3\sqrt{3}D/J + U_0$, $3 + v_0 - \sqrt{27(D/J)^2 + v_0^2 - 6\sqrt{3}DU_0/J + U_0^2}$ and $3 + v_0 + \sqrt{27(D/J)^2 + v_0^2 - 6\sqrt{3}DU_0/J + U_0^2}$. With these eigenvalues we can determine the 2D manifolds in the $(v_0, U_0, D)$ space which close the band gaps at $\pm K$ valleys. The manifolds are plotted in Fig.2 within a region $\mathcal{R}$ of the parametric space extended over the intervals $0 \leq v_0 \leq 0.5$, $0 \leq D \leq 0.15J$, and $0 \leq U_0 \leq 0.3$. The purple, green and blue surfaces all correspond to band gap closure at the $K$ valleys. Points on the red surface close the band gap at the $-K$ valleys. One can safely claim that the chosen parametric region $\mathcal{R}$ in the $(v_0, U_0, D)$ space covers all Van Der Waals ferromagnets with weak DMI and hence collinear magnetic order. Note that extending the DMI range to $0.3J$ (or even beyond) does not yield any new results.

The band closure manifolds divide $\mathcal{R}$ into six subregions denoted by Roman numerals I to VI in Fig.2. The cyan curve (called $\zeta$) represents an oriented elliptic arc penetrating all six regions and defined by $(v_0, U_0, D/J) = (0.17, 0.25\sin(u), 0.12\cos(u))$ with $0 \leq u \leq \pi/2$. In Fig.3a, we plot the minimal gaps between bands $\in_i$ and $\in_j$, denoted $g_{ij}$, along the arc $\zeta$. In consistency with Fig. 2, one of these minimal gaps vanishes when $\zeta$ intersects a band closure manifold. In particular, $g_{21}$ vanishes at the $K$ valley when $\zeta$ intersects the blue ($u \approx 0.3$) and purple ($u \approx 1.47$) surfaces; this is further illustrated in the magnon bands calculation presented in Figs. 3c and 3g respectively. In its turn, $g_{34}$ vanishes at the $-K$ valley when $\zeta$ intersects twice the red surface at $u \approx 0.41$ and $u \approx 1.36$, which is again confirmed in Figs. 3d and 3f respectively. Finally, the primary gap $g_{23}$ vanishes only once (Fig. 3a) when $\zeta$ intersects the green surface near $u \approx 1.19$ (Fig.2), closing the gap between $\in_2$ and $\in_3$ at the $K$ valley (Fig.3e).

Another interesting observation in Fig.2 is the intersection between the green and red surfaces along a straight line $\mathcal{L}$ given parametrically by $(v_0, U_0, D/J) = (3\sqrt{3}\,u, 3\sqrt{3}\,u, u)$. The minimal band gaps $g_{23}$ and $g_{34}$ vanish along $\mathcal{L}$ for any value of $u = D/J$ (Fig.3b), which closes the gaps at $-K$ ($g_{23} = 0$) and $K$ ($g_{34} = 0$) simultaneously. The band gaps closure is further illustrated in Fig.3h for the choice $D = 0.05J$.

Next, the Berry curvatures and Chern numbers analysis proves that the regions I to V correspond to five distinct topological phases with Chern numbers $[0, -2, 0, 2]$, $[0, -2, 1, 1]$, $[-1, -1, 1, 1]$, $[-1, 1, -1, 1]$, and $[0, 0, -1, 1]$ respectively. Region VI, however, corresponds to a topologically trivial phase with zero Chern numbers. The Hamiltonian hence evolves non-adiabatically along the arc $\zeta$ or any other curve that penetrates a gap closure manifold.

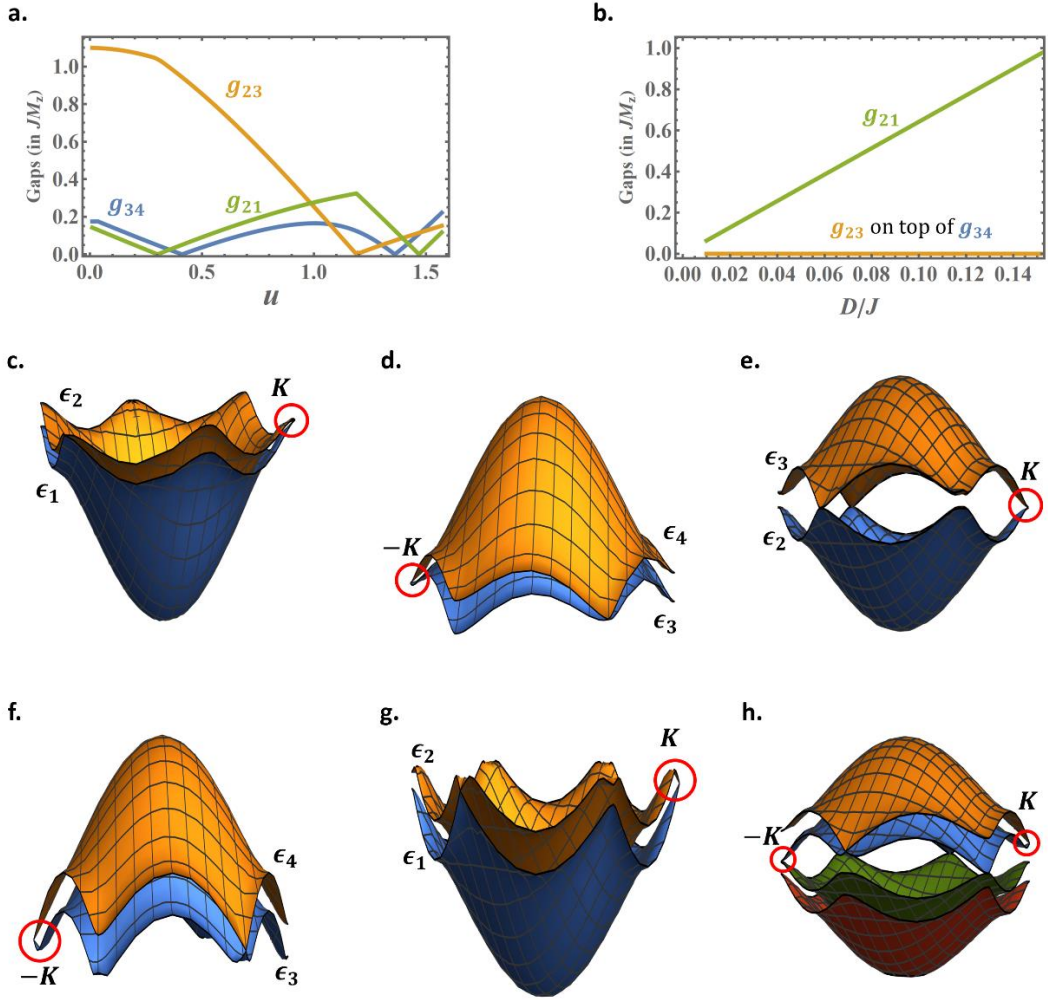

**Figure 3**: (a) Plot of the minimal gaps $g_{ij}$ between bands $\epsilon_i$ and $\epsilon_j$ along the elliptic arc in Fig.2. The arc is parametrized as $(v_0, U_0, D/J) = (0.17, 0.25\sin(u), 0.12\cos(u))$ with $0 \leq u \leq \pi/2$. (b) Plot of $g_{ij}$ along the line of intersection of the red and green manifolds in Fig.2. (c)-(d) Numerically calculated magnon bands at the gap closure points in (a), with $u \approx 0.3, 0.41, 1.19, 1.36$, and $1.47$ respectively. These correspond to the intersection between the elliptic arc and the band closure manifolds in Fig.2.

We have also studied the topological response of the HBF in terms of the thermal Hall conductivity ($\kappa_{xy}$). Our numerical investigation proves a strong dependence of $\kappa_{xy}$ on the DMI, regardless of the topological phase. As a general conclusion, $\kappa_{xy}$ is found to significantly increase (in absolute value) with the DMI. A representative sample of our numerical results is presented in Fig. 4a. On the contrary, $\kappa_{xy}$ is found to vary slightly when changing the value of the electrostatic potential $U_0$ or the interlayer exchange $v_0$ (Fig.4b). More important, we calculate in Fig.4c the saturation value of $\kappa_{xy}$ along $\zeta$ as it penetrates the five distinct topological phases. In this calculation, we excluded the topologically trivial phase VI and the gap closure points on $\zeta$. The profile of $\kappa_{xy}$ along $\zeta$ is found to be continuous and smooth, without abrupt jumps accompanying the nonadiabatic evolution. The results further confirm our previously stated conclusion concerning the dependence of the $\kappa_{xy}$ on the DMI; both $D$ and $\kappa_{xy}$ decrease with increasing values of the parameter $u$.

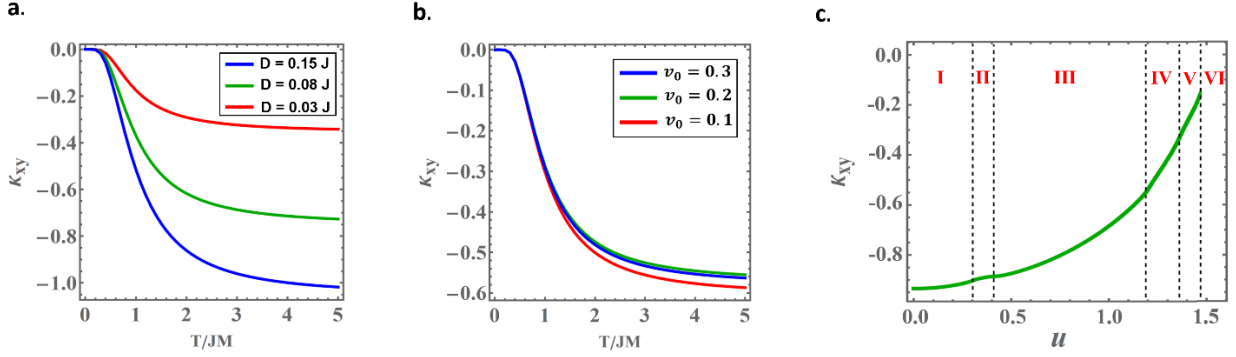

**Figure 4**: (a) Magnon thermal Hall effect ($\kappa_{xy}$) as a function of the temperature for selected values of $D/J$ with $(U_0, v_0) = (0.2, 0.3)$. (b) Same as (a) but for selected values of $v_0$, with $(U_0, D) = (0.1, 0.05J)$. Figure (b) illustrates the weak dependence of $\kappa_{xy}$ on $v_0$ and a similar conclusion holds for $U_0$. (c) Plot of the saturation value of $\kappa_{xy}$ on the elliptic curve $\zeta$ penetrating the topological regions I, ..., V. The dashed lines highlight the boundaries between the regions. The plot is truncated at the boundary between region V and the topologically trivial region VI.

## IV. Topological edge states in bilayer nanoribbons

The rich topology in the presented model motivates the investigation of edge magnons in its nanoribbon version. We consider a bilayer AB-stacked nanoribbon with zigzag edges. The ribbon is infinite in the $x-$direction and has a finite width in the $y-$direction. Fig. 1, truncated at the right and left edges, can serve as a schematic representation to illustrate the geometry. Each layer is composed of $N$ sites along the $y-$direction, denoted by $m = 1, \dots, N$ from right to left. The semi-classical approach in Section II can be used to derive the Schrödinger equation and momentum-space Hamiltonian for the nanoribbon. In the present case, the translation invariance is preserved along $x$ in both layers and allows for partial Fourier transform only along this direction. The Hamiltonian of each layer consequently reduces to an ensemble of $N$ one-dimensional lattice Hamiltonians, indexed by $k_x$. The $2N \times 2N$ Hamiltonian for the bilayer nanoribbon, acting on the direct sum of the Hilbert spaces of the layers, can then be constructed as

$$\mathcal{H}(k_x) = |1\rangle\langle 1| \otimes \mathcal{H}_1(k_x) + |2\rangle\langle 2| \otimes \mathcal{H}_2(k_x) + [|1\rangle\langle 2| \otimes \mathcal{H}_{12}(k_x) + h.c.]$$
(9a)

In Eq.9a, the symbol $\otimes$ denotes tensor product, h. c. stands for Hermitian conjugate and vectors $|l = 1, 2\rangle$ account for the layer degree of freedom. The intralayer and interlayer Hamiltonians, $\mathcal{H}_l$ and $\mathcal{H}_{12}$ respectively, read

$$\mathcal{H}_1 = M_z[2J + U + 2D \sin(k_x a)]|1\rangle\langle 1| + M_z[2J + J_\perp + U - 2D \sin(k_x a)]|N\rangle\langle N|$$

$$+ \sum_{m=2}^{N-1} G_{m,+}|m\rangle\langle m| + J_\perp M_z \sum_{m=1}^{N/2} |2m\rangle\langle 2m| - JM_z \sum_{m=1}^{N/2-1} [|2m\rangle\langle 2m+1| + h.c.]$$

$$+ F_1 \sum_{m=1}^{\frac{N}{2}} [|2m-1\rangle\langle 2m| + h.c.] + F_2 \sum_{m=1}^{N-2} [(-1)^m |m\rangle\langle m+2| + h.c.]$$

(9b)

$$\mathcal{H}_2 = M_z[2J + J_\perp - U + 2D \sin(k_x a)]|1\rangle\langle 1| + M_z[2J - U - 2D \sin(k_x a)]|N\rangle\langle N|$$

$$+ \sum_{m=2}^{N-1} G_{m,-}|m\rangle\langle m| + J_\perp M_z \sum_{m=1}^{N/2} |2m-1\rangle\langle 2m-1| - JM_z \sum_{m=1}^{N/2-1} [|2m\rangle\langle 2m+1| + h.c.]$$

$$+ F_1 \sum_{m=1}^{\frac{N}{2}} [|2m-1\rangle\langle 2m| + h.c.] + F_2 \sum_{m=1}^{N-2} [(-1)^m |m\rangle\langle m+2| + h.c.]$$

(9c)

To simplify the expressions, we have defined the functions

$G_{m,\pm}(k_x) = 3JM_z + 2(-1)^{m-1}DM_z \sin(k_x a) \pm U$, $F_1(k_x) = -2JM_z \cos(ak_x/2)$ and $F_2(k_x) = 2DM_z \sin(ak_x/2)$

The eigenvalues of $\mathcal{H}(k_x)$ yield the bulk and edge spectra for the nanoribbon. Bulk magnons are delocalized along $x$ and $y$ directions, while intra-gap edge states are exponentially confined to the edges, forming quasi-1D channels. Figures 5a to 5e present numerical results for nanoribbons with $N = 100$, in the topological phases I to V respectively. We consider a large $N$ which (almost) reproduces the bulk bands (blue modes) of the infinite bilayer. Topological edge modes are observed in the band gaps, near $\pm K$ valleys, in agreement with the bulk-edge correspondence.

Numerical investigation verified the robustness of the edge modes against the edge values of the DMI, exchange and ED parameters. Interestingly, the bulk and edge modes in Figs.5a-5e are nonreciprocal ($\epsilon_n(-k_x) \neq \epsilon_n(k_x)$) as a consequence of the ED (without ED, the modes are indeed reciprocal [17]). Bulk and edge magnons hence propagate with different energies in opposite directions. For completeness, we present in Fig. 5f the bulk and edge modes in the topologically trivial phase VI.

The color code for the edge modes indicates their localizations. The red and magenta modes are respectively localized on the right and left edges of layer 1. Bright and dark green modes are confined to right and left edges of layer 2 respectively. An edge mode hence picks a specific channel and cannot propagate on more than one edge simultaneously. Moreover, the group velocity (slope of the dispersion curve) shows that modes on right and left edges propagate in opposite directions. The topological nature, nonreciprocity and asymmetric propagation of these edge modes might be interesting for applications in magnonic devices.

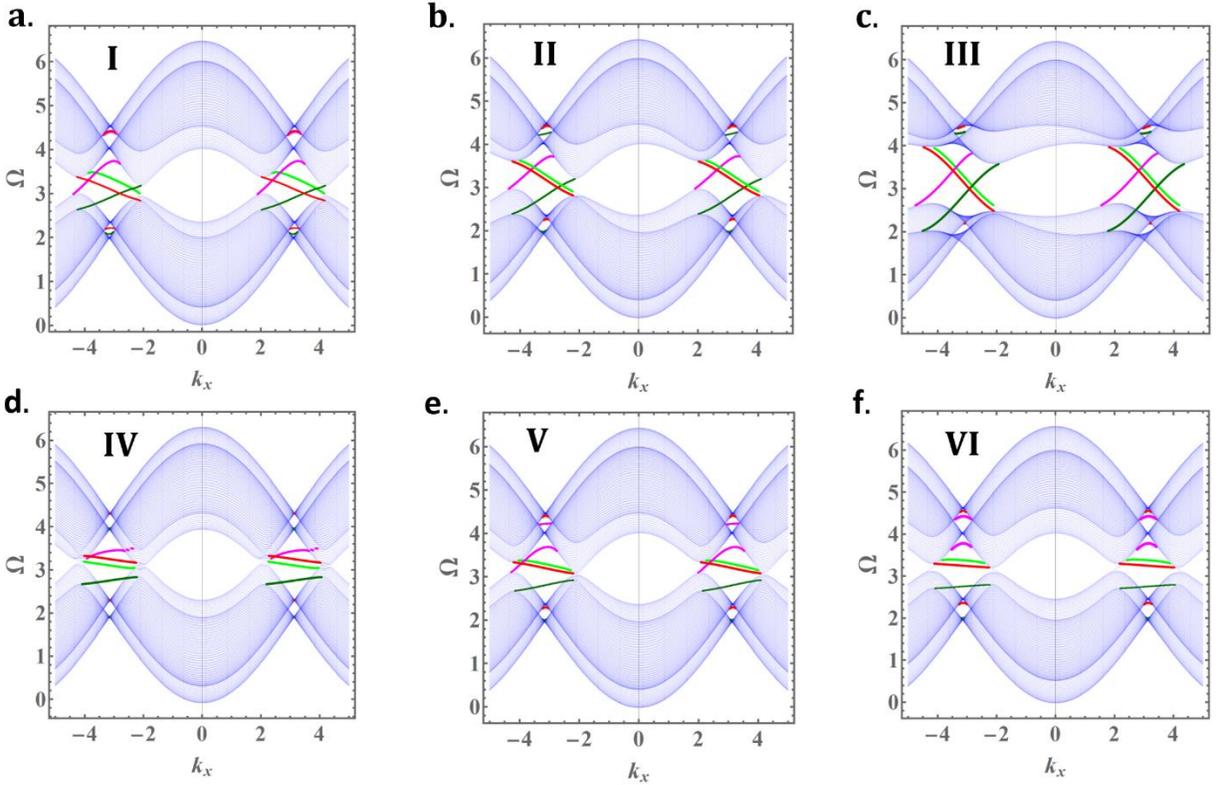

**Figure 5**: Bulk and intra-gap edge modes for a bilayer nanoribbon with $N = 100$ and zigzag edges. Figures (a)-(d) correspond to topological phases I-IV respectively. Figures (e) and (f) correspond to ED-free and DMI-free nanoribbons respectively. The parameters are (a) $(U_0, v_0, D) = (0.1, 0.4, 0.05J)$, (b) $(U_0, v_0, D) = (0.2, 0.3, 0.08J)$, (c) $(U_0, v_0, D) = (0.2, 0.3, 0.15J)$, (d) $(U_0, v_0, D) = (0.2, 0.3, 0.025J)$, (e) $(U_0, v_0, D) = (0, 0.4, 0.1J)$, and (f) $(U_0, v_0, D) = (0.2, 0.3, 0)$.

## V. Conclusion

Similar to the potentials of bilayer graphene in nanoelectronics, bilayers formed of 2D magnetic materials shall offer promising opportunities for magnonic and spintronic nano-devices. In particular, electrostatic doping in magnetic bilayers is gaining increasing attention as an important contributor to the exotic physics underlying their magnetic excitations. Compared to their fermionic counterpart, however, magnetic bilayers remain less explored which motivated our study. A generalized model for HBF with both ED and DMI is presented and the important topological consequences of simultaneously breaking inversion symmetry (by ED) and time reversal symmetry (by DMI) are highlighted. The model turns out to be topologically rich and worth attention from fundamental and applied perspectives. We predicted five distinct topological phases and studied in details the corresponding edge spectra, Hall conductivities, and nonadiabatic evolution in the parameter space of the four-band Hamiltonian. The presented parametric study is general and covers all HBFs with collinear order. We concluded that the nonadiabatic evolution can be induced by varying any of the Hamiltonian parameters (interlayer exchange, DMI or ED). Worth noting that ED and interlayer exchange can be experimentally tuned in bilayer systems [30, 67] and our analysis offers a route to manipulate Chern numbers and topological response in HBF. Further, the results promote ED as a promising experimental technique to enrich the magnon physics in 2D bilayers. In addition to its effect on the topology, ED is found to break the reciprocity of bulk and edge states which is desired for technological applications. It should be interesting to extend the present investigation towards other possible ground states (e.g. layer antiferromagnets [14]) and bilayer stacking (e.g. twisted bilayers [42, 49, 68, 69]).